\documentclass[journal=nalefd,manuscript=letter]{achemso}

\usepackage[T1]{fontenc} 
\usepackage[normalem]{ulem}
\usepackage{amsmath,amssymb}
\usepackage{multirow}
\usepackage{xcolor,soul}
\usepackage{srcltx}
\usepackage{hyperref,graphicx}
\usepackage{siunitx}

\usepackage{chemformula} 
\usepackage[T1]{fontenc} 

\usepackage{etoolbox}

\makeatletter

\newcommand\tcr{\textcolor{black}}

\renewcommand\st[1]{}

\author{Mert Akturk}
\affiliation[Polimi]{ Dipartimento di Fisica - Politecnico di Milano, Piazza Leonardo da Vinci, 32, I-20133 Milano, Italy}
\altaffiliation{These authors contributed equally}

\author{Giulia Crotti}
\affiliation[Polimi]
{ Dipartimento di Fisica - Politecnico di Milano, Piazza Leonardo da Vinci, 32, I-20133 Milano, Italy}
\altaffiliation{These authors contributed equally}

\author{Andrea Schirato}
\affiliation[Polimi]{ Dipartimento di Fisica - Politecnico di Milano, Piazza Leonardo da Vinci, 32, I-20133 Milano, Italy}
\alsoaffiliation[Rice]{ Department of Physics and Astronomy, Rice University, Houston, Texas 77005, United States}

\author{Katsuya Tanaka}
\affiliation[UniJeSSP]{Institute of Solid State Physics, Friedrich Schiller University Jena, Jena 07743, Germany}
\alsoaffiliation[UniJe]{ Institute of Applied Physics, Abbe Center of Photonics, Friedrich Schiller University, Jena, Jena, 07745, Germany}

\author{Isabelle Staude}
\affiliation[UniJeSSP]{Institute of Solid State Physics, Friedrich Schiller University Jena, Jena 07743, Germany}
\alsoaffiliation[UniJe]{ Institute of Applied Physics, Abbe Center of Photonics, Friedrich Schiller University, Jena, Jena, 07745, Germany}

\author{Thomas Pertsch}
\affiliation[UniJe]{ Institute of Applied Physics, Abbe Center of Photonics, Friedrich Schiller University, Jena, Jena, 07745, Germany}

\author{Giulio Cerullo}
\affiliation[Polimi]{ Dipartimento di Fisica - Politecnico di Milano, Piazza Leonardo da Vinci, 32, I-20133 Milano, Italy}

\author{Giuseppe Della Valle}
\email{giuseppe.dellavalle@polimi.it}
\affiliation[Polimi]{ Dipartimento di Fisica - Politecnico di Milano, Piazza Leonardo da Vinci, 32, I-20133 Milano, Italy}
\altaffiliation{These authors jointly supervised this work}

\author{Margherita Maiuri}
\email{margherita.maiuri@polimi.it}
\affiliation[Polimi]{ Dipartimento di Fisica - Politecnico di Milano, Piazza Leonardo da Vinci, 32, I-20133 Milano, Italy}
\altaffiliation{These authors jointly supervised this work}

\title{Spatiotemporal Metasurface for Ultrafast All-Optical Wavefront Shaping }

\keywords{Ultrafast nanophotonics, Optical metasurfaces, Wavefront shaping, Ultrafast imaging}

\begin{document}


\newpage
\begin{abstract} 
    Optical metasurfaces have established themselves as exceptional nanophotonic platforms to control the properties of light at subwavelength scales. 
    By properly designing the geometry and spatial arrangement of their constituent meta-atoms, engineered phase gradients can be imparted to an incoming field, thereby tailoring its wavefront with a virtually unlimited number of control knobs. 
    However, once fabricated, a metasurface has a permanently fixed optical response. 
    Here, we introduce a strategy for dynamic control, by theoretically predicting and experimentally demonstrating all-optical, reversible, and ultrafast wavefront shaping in a periodic semiconductor nanowire metasurface. 
    A strongly astigmatic femtosecond optical pump pulse is used to photoexcite the resonant metasurface non-uniformly, inducing a transient, spatially inhomogeneous permittivity modulation that reshapes the wavefront of a delayed probe pulse. Ultrafast imaging measurements reveal transient defocusing of the transmitted field at selected probe wavelengths, with a switching time below one picosecond. 
    Quantitative numerical modelling elucidates the defocusing mechanism as arising from the interplay between the unperturbed optical response of the resonant metasurface and the pump-induced spatial permittivity gradient. 
    Our work establishes a fully all-optical and ultrafast route to the dynamic analogue of passive wavefront shaping in gradient metasurfaces, paving the way to all-optically reconfigurable metalenses.

\end{abstract}

\newpage
\section{Introduction}

Optical metasurfaces have emerged as a unique platform for manipulating light at the nanoscale. 
These quasi-two-dimensional artificial materials, composed of periodically arranged sub-wavelength nanoresonators, enable complex optical functionalities within structures only a few hundred nanometers thick \cite{yu2011,yu2014,kildishev2013,Kuznetsov2024,Schulz2024,brongersma2025}, thereby overcoming many limitations of conventional optical components \cite{arbabi2015,neshev2018}.
At the core of their operation lies the rational design and engineering of the geometry, composition, and arrangement of the constituent nanoantennas (commonly referred to as `meta-atoms'), which collectively determine the optical response of the metasurface, and enable unprecedented control over the amplitude, phase, polarisation, and angular momentum of light \cite{overvig2019,ren2019,chen2016,quadruplex2020}.\\

One of the most successful applications of metasurfaces is wavefront shaping. 
In particular, metalenses are flat-optical elements in which the spatially varying response of individual nanoresonators is engineered to impart a specific phase profile to an incident electromagnetic field. 
The desired wavefront transformation is achieved by encoding the required phase profile in the geometry of the constituent meta-atoms. Different mechanisms have been exploited to control the phase of light in metalenses, including the geometric (or Pancharatnam-Berry) phase, the propagation phase, and the resonance phase. 
In most implementations, the targeted phase profile is obtained through a lookup-table approach, where the geometrical parameters of the meta-atoms are varied to provide the required phase delay while preserving high transmission. 
Regardless of the specific phase mechanism, the optical response is entirely determined by the geometry and spatial arrangement of the meta-atoms, which is set at the design stage. This approach provides powerful wavefront control, but results in a static optical functionality once the structure is fabricated. \\

This intrinsic limitation has motivated the development of dynamic metasurfaces, in which the optical response experienced by a signal beam can be reversibly reconfigured even after fabrication. 
Several reconfiguration mechanisms have been explored, including thermal, mechanical, electrical and phase-change approaches, providing modulation speeds that span several orders of magnitude \cite{shaltout2019,reconfig2019}.
Among these, all-optical control allows the fastest operation \cite{vasa2009, makarov2017, taghinejad2019, carletti2021,vabishchevich2021, hail2026}, as it exploits photoinduced optical nonlinearities typically by `hot' carriers, i.e.~nonequilibrium electronic states with high (up to a few eV) excess energy and ultrashort (femtoseconds to picoseconds) lifetime \cite{Shcherbakov2015,Shcherbakov2017,DellaValle2017,Schirato2023,Tognazzi2023}. 
In an all-optical implementation, a first ultrashort pump pulse drives these transient nonlinearities, while a second, delayed probe pulse interrogates the modified optical response. As a result, optical modulations can occur on femto- to picosecond timescales, enabling reconfiguration rates up to hundreds of GHz, well beyond the capabilities of other dynamic tuning mechanisms  \cite{Maiuri2024,iyer2023,Shilkin_NanoLett_2024,crotti2024,Shilkin_NanoLett_2026}.\\

Since the advent of dynamic metasurfaces, a broad range of reconfigurable optical functionalities has been demonstrated, including wavefront shaping and light focusing.
Dynamic metalenses have been realized in the visible and near-infrared region \cite{kim2022,she2018, colburn2018}, using thermal tuning \cite{malek2024,archetti2022}, mechanical stretching \cite{afridi2023,ee2016,kamali2016}, electrical gating \cite{arbabi2018,bosch2021}, and all-optical excitation \cite{He,Shalaginov2021}. 

Despite these advances, most dynamic wavefront-shaping approaches remain based on a phase profile that is already permanently encoded in the fabricated metasurface. The external stimulus therefore changes the intensity, spectral position or focal length of a predesigned optical functionality, rather than generating the spatial phase gradient required for wavefront control. \cite{palermo2022alloptical,sinelnik2024ultrafast,Fan_ACSP_2023}. 
An alternative paradigm would be to illuminate an otherwise uniform metasurface with a spatially structured pump beam. 
In this approach, the pump beam does not merely tune an existing function, but transiently imprints a non-uniform permittivity landscape, that acts as a wavefront-shaping element for a delayed probe pulse.
Realizing this concept would provide a direct bridge between ultrafast all-optical modulation and flat-optical wavefront engineering, enabling optical functionalities that are imprinted on demand rather than permanently defined during fabrication. \\

In this work, we theoretically design and experimentally demonstrate ultrafast photoinduced wavefront shaping enabled by a spatiotemporally modulated metasurface. Rather than relying on a phase profile permanently encoded during fabrication, our approach generates a transient optical phase gradient through spatially structured photoexcitation. 
As a proof of concept, we demonstrate our approach in a one-dimensional periodic array of hydrogenated amorphous Silicon (a-Si:H) nanowires excited by a strongly astigmatic ultrashort pump pulse, which provides a simple implementation of structured optical excitation. 
By combining ultrafast imaging experiments with full-wave non-equilibrium simulations, we observe a transient sub-picosecond defocusing of the transmitted probe beam. 
We rationalize the effect in terms of a photoinduced inhomogeneous phase profile across the metasurface plane imprinted by the Gaussian spatial intensity profile of the pump pulse. 
The effect occurs in a spectral region where the metasurface response enhances phase modulation while minimizing amplitude changes, in close analogy with the operating principle of a static metalens.  
Taken together, these results establish a general  strategy for ultrafast, all-optically reconfigurable wavefront shaping based on photoinduced-gradient metasurfaces, in which spatially structured excitation and ultrafast optical nonlinearities are combined to create optical functionalities on demand, paving the way to high-speed, light-driven flat optical components.

\section{Results}
\label{sec:results}

\begin{figure}[h!]
    \centering
    \includegraphics[width=\textwidth]{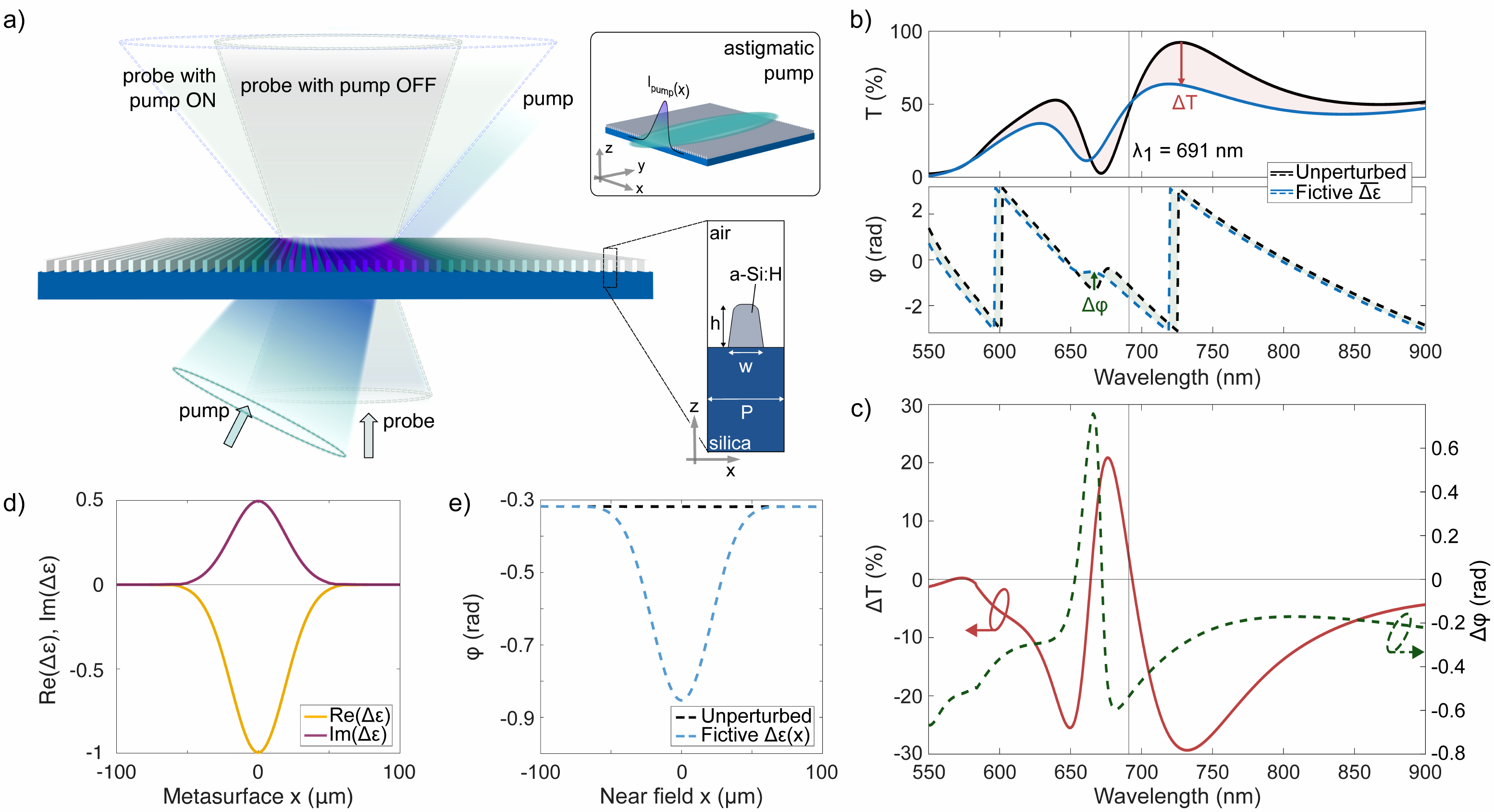}
    \caption{Design of a photoinduced dynamic metalens. (a), Schematic representation of inhomogeneous pump illumination and its effect on the probe wavefront. Bottom inset: Cross-section ($xz$-plane) of the metasurface unit cell, consisting of an a-Si:H nanowire on a silica substrate. The geometrical parameters are $P = \SI{400}{nm}$, $w = \SI{180}{nm} $, $h = \SI{220}{nm}$. Top inset: sketch of the pump excitation. (b), Simulated transmission intensity $T$ (solid curves, top) and phase $\varphi$ (dashed curves, bottom) of the metasurface under unperturbed conditions (black curves) or by imposing a fictive, spatially uniform permittivity variation $\overline{\Delta \varepsilon} = -1+\mathrm{i}0.5$ (blue curves). The illuminating plane wave is considered to be at normal incidence and polarized along the wires ($y$-axis). The vertical line corresponds to the wavelength $\lambda_1 = \SI{691}{nm}$ analysed in the following. (c), Differential intensity $\Delta T$ (red solid curve, left axis) and phase $\Delta \varphi$ (green dashed curve, right axis) between the perturbed and unperturbed conditions [see arrows in panel (a)]. (d), Spatially inhomogeneous permittivity perturbation $\Delta\varepsilon(x)$ (yellow for real part, purple for imaginary part) as a function of the $x-$coordinate across the metasurface plane. The peak value at $x = \SI{0}{\micro \m}$ was adjusted to coincide with the $\overline{\Delta \varepsilon}$ set in panels (b-c). (e) Spatial profile of the phase $\varphi$ of the transmitted probe field at $\lambda_1$, evaluated $\SI{2}{\micro\m}$ above the metasurface plane. The unperturbed case (black curve), yielding a flat spatial profile, is compared to the perturbed conditions (blue curve) where the $\Delta\varepsilon$ illustrated in panel (d) is applied, producing a Gaussian spatial profile.}
    \label{fig1}
\end{figure}

The operating principle of the spatiotemporally modulated metasurface is schematically illustrated in Fig.~\ref{fig1}a.
The structure consists of a 1D periodic array of identical a-Si:H nanowires on a silica substrate. 
A detailed cross-sectional view of the unit cell, together with the relevant geometrical parameters used for the metasurface design, is shown in the bottom inset of Fig.~\ref{fig1}a.
Since the nanowires are identical and periodically arranged along the $x-$direction, every unit cell is optically equivalent under unperturbed conditions.
As a result, the metasurface does not impart any spatial phase gradient to the incident beam. A normally incident probe beam therefore experiences a spatially uniform phase delay upon transmission, preserving its planar wavefront.\\ 

We dynamically break the symmetry by using light itself as a reconfigurable, non-lithographic spatial degree of freedom. To this end, we implement a pump-probe scheme in which a spatially structured photoexcitation transiently writes a phase gradient across the metasurface.
Specifically, a strongly astigmatic ultrashort pump pulse with a Gaussian intensity profile $I_{\text{pump}}(x)$ (blue beam in Fig.~\ref{fig1}a) excites the array, producing a highly non-uniform excitation along the periodic $x-$direction with a Gaussian-like intensity profile, while remaining nearly uniform in the longitudinal $y-$direction of the nanowires (see top right inset, Fig.~\ref{fig1}a).
Upon photoexcitation, hot electron-hole pairs mediate strong optical nonlinearities, transiently modifying the complex-valued permittivity of the material.
Therefore, a spatially varying pump absorption, following the pump intensity profile, is expected to generate a corresponding transient permittivity modulation $\Delta \varepsilon(x,t)$ across the metasurface plane, as illustrated by the colour gradient across the nanowires in Fig.~\ref{fig1}a.\\

As a result, a probe beam propagating through the photoexcited metasurface experiences a spatially varying phase delay that is reminiscent of the Gaussian-like pump intensity profile $I_{\text{pump}}(x)$.
At first order, we expect the system to operate as a transient cylindrical lens, producing either focusing or defocusing of the transmitted probe beam, depending on the concavity of the induced phase profile.
The defocusing case is schematically shown by the comparison between the pump-OFF and pump-ON probe cross-section profiles (gray beams) in Fig.~\ref{fig1}a.\\

However, photoinduced changes in the permittivity of the meta-atom typically modify the complex-valued transmission coefficient of the system, affecting both its amplitude and phase. 
Thus, in direct analogy with a static metalens, the desired operating condition for wavefront shaping occurs when amplitude variations are minimised, while phase modulation is enhanced. 
Therefore, the first step of our design strategy is to identify the spectral region in which the resonant metasurface converts a material perturbation predominantly into a phase shift, rather than amplitude modulation.
To this end, we numerically analysed the spectral response of the periodic metasurface, by comparing the transmitted intensity (squared amplitude) and phase, $T$ and $\varphi$, under unperturbed and out-of-equilibrium conditions.
Full-wave numerical simulations were performed for an infinite periodic array under unperturbed conditions and after imposing a spatially homogeneous fictitious permittivity variation $\overline{\Delta\varepsilon} = -1 + \mathrm{i}0.5$, representative of highly photoexcited Si nanostructures (on the order of few $\unit{m\joule\per\cm\squared}$) \cite{DellaValle2017}. 
Details on the simulations are provided in the Methods Section 2.1.

The main results of this design step are reported in Figs.~\ref{fig1}b-\ref{fig1}c.
Fig.~\ref{fig1}b shows the calculated transmission intensity $T$ (solid lines) and phase $\varphi$ (dashed lines) of the a-Si:H metasurface under unperturbed conditions (black traces) and after applying the spectrally and spatially homogeneous permittivity perturbation $\overline{\Delta \varepsilon}$  (blue traces). 
The unperturbed spectra show a resonance around $\SI{670}{nm}$, manifesting as a dip in intensity and a corresponding kink-like shape in the phase. 
At other wavelengths the transmission is characterized by broader spectral features. 
When the perturbation is introduced, all the features, including the main resonant dip, blue shift, consistently with the negative sign of $\text{Re}(\overline{\Delta\varepsilon})$, and broaden, due to the additional losses introduced by $\text{Im}(\overline{\Delta\varepsilon})$. \\

From the calculated spectra, we extracted the corresponding differential intensity and phase, $\Delta T$ and $\Delta\varphi$, shown in Fig.~\ref{fig1}c and defined as the difference between perturbed and unperturbed quantities (as marked respectively by the red and green arrows in Fig.~\ref{fig1}b, and highlighted by the shaded areas).
These differential spectra are a key part of our design strategy.
In close analogy to the conventional lookup-table approach for static metalenses, where the relationship between the geometry of the meta-atom and the resulting transmission amplitude and phase is established before the desired wavefront is assembled, here we map how a photoinduced material perturbation is converted by the resonant metasurface into $\Delta T$ and $\Delta \varphi$.
As shown in Fig.~\ref{fig1}c, both quantities exhibit pronounced wavelength-dependent variations in both magnitude and sign, reflecting the spectral dispersion of the metasurface, even if the $\overline{\Delta\varepsilon}$ introduced here is spectrally uniform. 

This is especially evident around the resonance, where both $\Delta T$ and $\Delta \varphi$ display sign changes and negative or positive extrema.
We focus in particular on the wavelength $\lambda_1 = \SI{691}{nm}$ (highlighted by the vertical line), located on the red wing of the resonance.
At this wavelength, $\Delta \varphi$ is close to a negative peak, while $\Delta T$ approaches zero (isosbestic point). 
Therefore, at $\lambda_1$ the optical perturbation predominantly affects the phase of the transmitted light, without significant intensity variations. 
In the sign convention adopted in our models, a negative value of $\Delta\varphi$ indicates a reduced phase retardation with respect to the unperturbed condition.
In analogy with propagation and phase accumulation through a conventional optical element, therefore, the photoexcited metasurface behaves as an optically thinner medium with respect to its unperturbed counterpart.\\

We then explicitly included the spatial dependence of the photoexcitation into the design.
We performed finite-aperture full-wave simulations of the metasurface under a representative Gaussian-shaped permittivity modulation $\Delta \varepsilon(x)$, imposed along the $x-$direction.
The purpose of this step is to demonstrate that, at the selected probe wavelength, a non-uniform material perturbation can be translated into a non-uniform phase profile across the metasurface plane, thereby providing the design basis for all-optically induced defocusing.

For this analysis, we considered a finite-sized metasurface with a total width of $\SI{300}{\micro m}$ along the $x$-direction comprising more than 730 meta-atoms, and illuminated with a collimated Gaussian probe beam whose waist was placed on the metasurface plane.
The probe intensity full width at half maximum (FWHM) was set to $I_{\text{probe}} = \SI{70}{\micro m}$, which is consistent with experimental conditions. 
Details on the model implementation are provided in Methods Section 2.2.
We compared the unperturbed system to a photoexcited configuration with a space-dependent nanowire permittivity  $\Delta \varepsilon(x)$ varying as a Gaussian function along the $x-$coordinate. 
Its FWHM is $\SI{45}{\micro m}$, in line with the envisioned FWHM value of the pump intensity $I_x$. 
This is shown in Fig.~\ref{fig1}d, where the real and imaginary parts of $\Delta\varepsilon(x)$ are reported as yellow and purple traces, respectively.
For a quantitatively consistent estimate, the peak modulation at $x = \SI{0}{\micro m}$ was chosen to match the spatially homogeneous $\overline{\Delta\varepsilon}$ used in Figs.~\ref{fig1}b-c. 

The resulting phase of the transmitted beam, extracted in the near field at a distance of $\SI{2}{\micro m}$ above the metasurface plane, is reported in Fig.~\ref{fig1}e as a function of the $x-$coordinate. 
In unperturbed conditions, the phase $\varphi$ remains essentially flat across the metasurface aperture, as expected for a collimated beam transmitted through a periodic array of nanowires with no built-in phase gradient along the direction of periodicity. 
Conversely, when the spatially inhomogeneous perturbation is applied (blue curve, Fig.~\ref{fig1}e), the transmitted probe phase acquires a Gaussian shape, directly reminiscent of $\Delta\varepsilon(x)$.
Its upward concavity is consistent with the negative sign of $\Delta \varphi$ identified in Fig.~\ref{fig1}c. 
The phase retardation is reduced most strongly at the centre of the metasurface, where the permittivity modulation is stronger, and progressively vanishes towards the tails of the pump beam.
The central region  of the photoexcited metasurface is therefore optically thinner than its edges, so that it behaves as a diverging cylindrical lens, producing beam defocusing. 
These results establish the design principle underlying our approach.\\

To experimentally validate the predicted photoinduced wavefront shaping, we performed ultrafast imaging experiments. 
The fabricated sample consists of a 1D array of a-Si:H stripes which, for the used probe polarisation, behave as nanowires with tapered rectangular cross-section.
A top view scanning electron microscopy (SEM) image of the sample is shown in the top inset of Fig.~\ref{fig3}a, while fabrication details are provided in the Supplementary Information (SI) Section \tcr{S1}. 
The pump-probe ultrafast imaging setup is schematically shown in Fig.~\ref{fig3}a. 
A strongly astigmatic pump pulse at $\lambda_\text{pump}=400$~nm, i.e.~above the Si bandgap, with a temporal duration of $\sim 100$ fs photoexcites the metasurface nonuniformly along the $x-$direction.
The pump pulse FWHMs were estimated as $I_x = \SI{45}{\micro m}$ and $I_y = \SI{4}{mm}$, in agreement with the numerical design.
A delayed probe pulse interrogates the excited region, and the transmitted probe beam is imaged onto a CMOS camera through relay optics with focal lengths $f_{\text{L1}} = \SI{75}{\milli\meter}$ and $f_{\text{L2}} = \SI{100}{\milli\meter}$. 
The probe is  spectrally filtered around \SI{690}{\nano\meter}, corresponding to the phase-dominated operating region identified above.
Further details of the experimental conditions are provided \tcr{in Methods Section 1}. 

Figs.~\ref{fig3}b--\ref{fig3}c show representative probe beam profile images recorded at two different pump-probe delays. 
Before photoexcitation, at $t = \SI{-1.4}{\pico\second}$ (Fig.~\ref{fig3}b), the probe exhibits an unperturbed Gaussian intensity profile with a FWHM of \SI{0.28}{\milli\meter} in the detection plane, as highlighted by the intensity contour. 
Following photoexcitation, at $t = \SI{0.1}{\pico\second}$ (Fig.~\ref{fig3}c), the transmitted probe broadens along the $x$-direction, reaching a FWHM of approximately \SI{0.44}{\milli\meter}.
The pronounced pump-induced broadening demonstrates a transient modification of the probe wavefront, in excellent agreement with the behaviour predicted by the design simulations, and provides direct evidence of ultrafast photoinduced defocusing.


\begin{figure}[ht!]
    \centering
    \includegraphics[width=.8\textwidth]{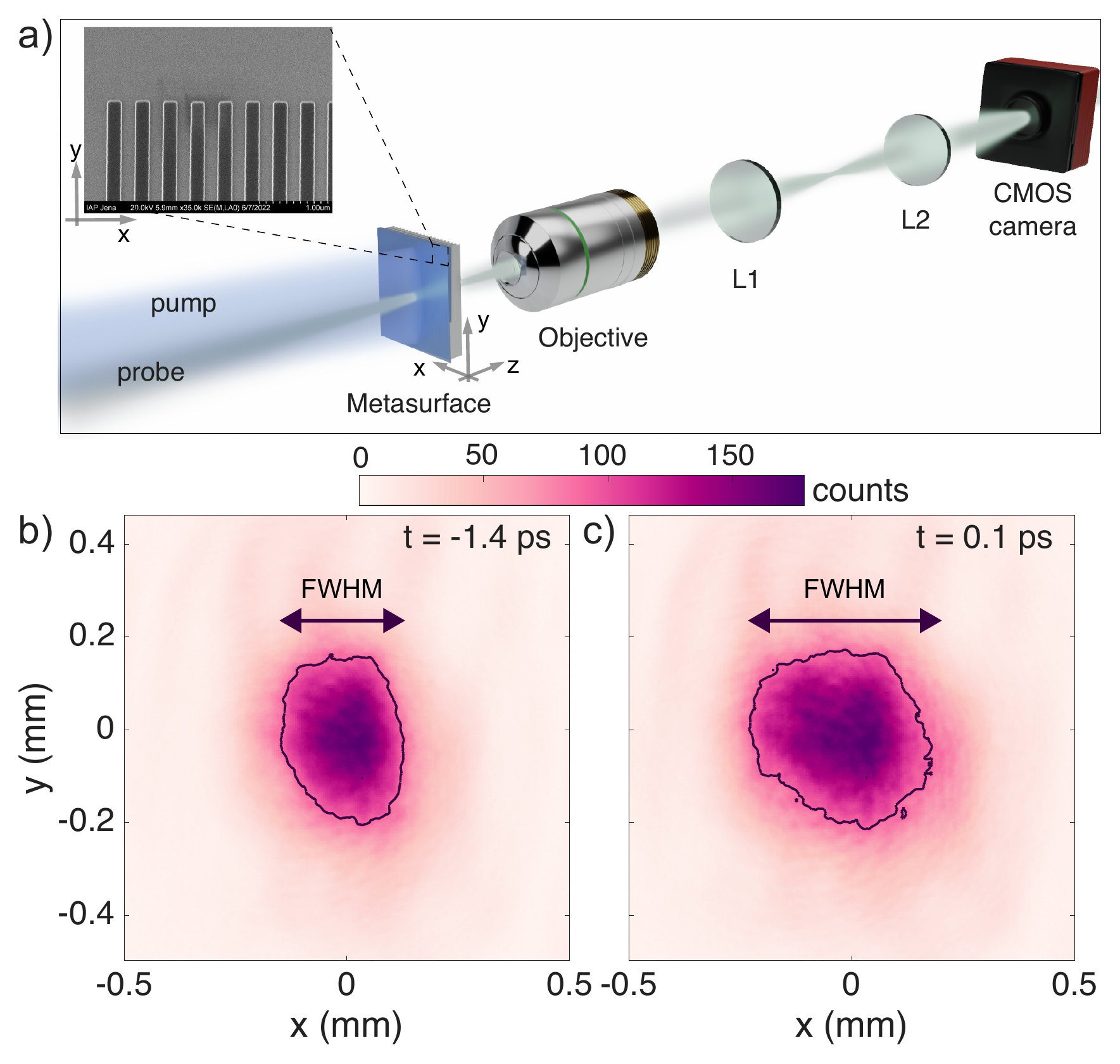}
    \caption{Experimental setup and intensity maps. 
    (a) Sketch of the ultrafast imaging setup. The inset shows a top-view SEM image of the sample (horizontal $xy$-plane). (b-c) Measured intensity maps captured by the CMOS camera at different pump-probe time delays, $\SI{-1.4}{ps}$ (b) i.e.~unperturbed system, and $\SI{0.1}{ps}$ (c). Line contours identify the corresponding intensity FWHM of the spot, estimated to $\text{FWHM} \simeq \SI{0.28}{mm}$ (b) and $\text{FWHM} \simeq \SI{0.44}{mm}$ (c), respectively.
    }
    \label{fig3}
\end{figure}

To quantify the transient wavefront modulation, we introduce a differential figure of merit. 
For a given pump-probe delay, $t$,  we first average the CMOS camera counts along the $y-$axis to obtain the transmitted probe intensity $I(x,t)$, recorded as a function of the $x-$coordinate.
We then use as reference the unperturbed transmitted probe intensity profile $I_0(x)$ measured at $\SI{-1.4}{ps}$. 
Finally, we define:
\begin{equation}
    \Delta I/I_0 (x,t) = \frac{I(x,t)-I_{0}(x)}{I_0(0)},
    \label{eq:DI_I}
\end{equation}
where the normalization constant $I_0(0)$ is the maximum of the unperturbed transmitted intensity, corresponding to the beam centre, in $x = \SI{0}{\micro m}$. 

\begin{figure}[ht!]
    \centering
    \includegraphics[width=\textwidth]{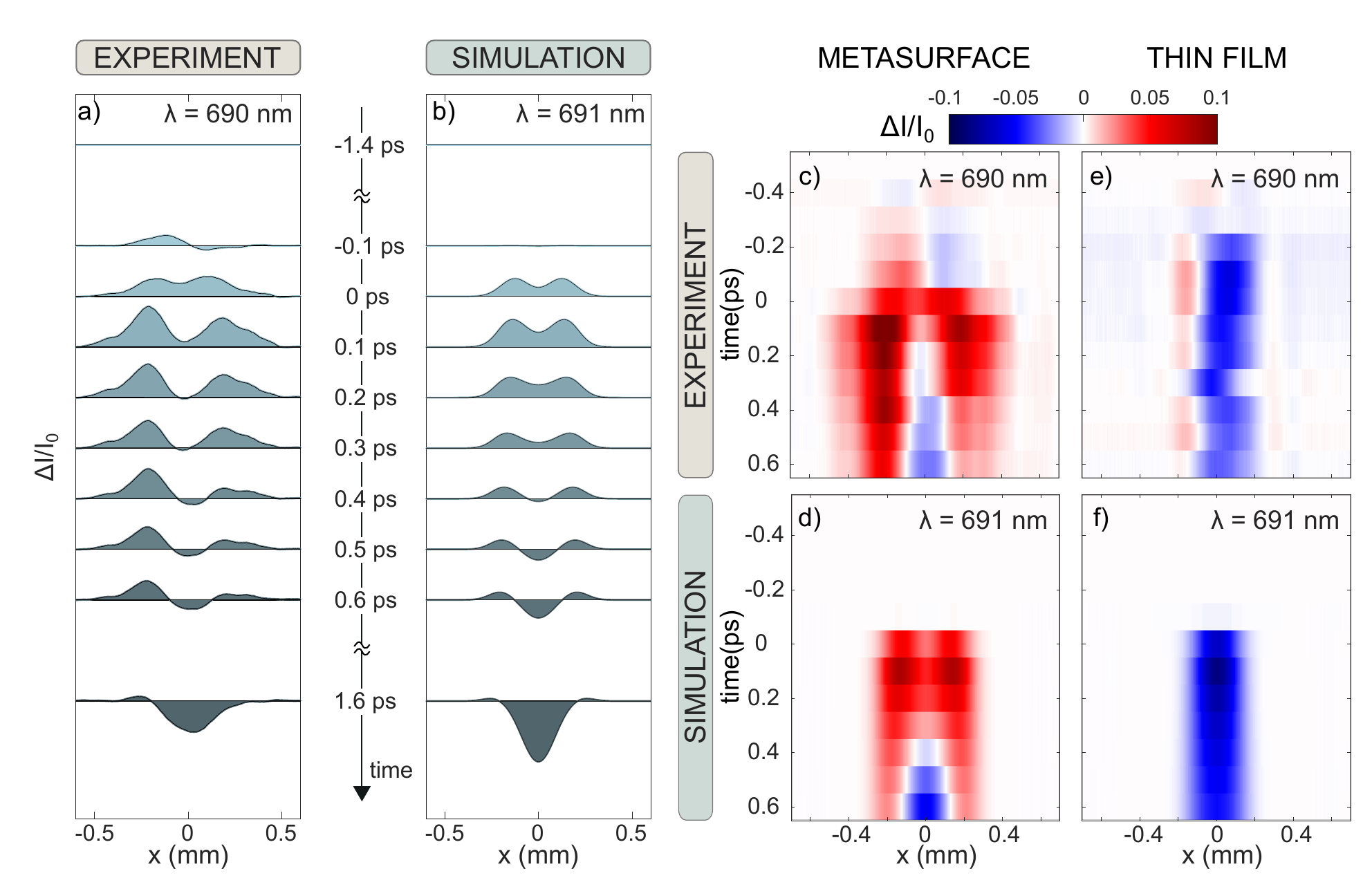}
    \caption{Ultrafast wavefront shaping.
    (a-b) Measured (a) and simulated (b) $\Delta I/I_0$ signals of the metasurface, at different pump-probe time delays, for probe wavelengths $\lambda = \SI{690}{nm}$ and $\lambda = \SI{691}{nm}$, respectively, as a function of the spatial $x-$coordinate on the CMOS camera. (c-d) $\Delta I/I_0$ experimental (c) and simulated (d) maps as a function of $x$-coordinate and pump-probe delay for the photoexcited metasurface. (e-f) Same as (c-d) for the control experiment performed on a thin film of comparable thickness at the same probe wavelength. 
    }
    \label{fig4}
\end{figure}

The measured $\Delta I/I_0$ profiles are shown in Fig.~\ref{fig4}a for different pump-probe delays. 
The signature of probe defocusing is a double-peaked differential profile, reflecting the pump-induced redistribution of intensity from the centre towards the lateral regions of the beam.
This distinctive feature appears within the temporal overlap of pump and probe pulses and decays on the sub-picosecond timescale, disappearing within approximately $\SI{600}{fs}$.
At longer delays the differential profile  evolves into a negative dip, as shown by the trace at $\SI{1.6}{ps}$ in Fig.~\ref{fig4}a.  We attribute this behaviour to the increasing dominance of photoinduced attenuation of the transmitted probe. Although the phase-induced defocusing effect persists, its contribution becomes largely obscured by the stronger amplitude modulation.  
The stacked $x-$dependent traces in Fig.~\ref{fig4}a are complemented by the 2D maps in Fig.~\ref{fig4}c, where $\Delta I/I_0$ is displayed as a function of both the $x-$coordinate on the CMOS camera (horizontal axis) and the pump-probe delay (vertical axis).\\

To elucidate the physical origin of the measured transient response, we developed a quantitative numerical model that reproduces the full ultrafast imaging experiment. 
Starting from the photoinduced nonequilibrium carrier dynamics a-Si:H, the model reconstructs the resulting spatiotemporal permittivity variations $\Delta \varepsilon(x,t)$, and finally computes how it reshapes the transmitted probe wavefront under the same finite-aperture conditions employed in the experiment. 

The photoinduced optical response of a-Si:H was modelled by accounting for the dominant nonequilibrium processes activated upon ultrafast photoexcitation \cite{DellaValle2017}. 
Absorption of the pump pulse, through both one- and two-photon absorption (TPA) processes, generates a population of out-of-equilibrium electron-hole pairs inside the nanowires. 
The subsequent relaxation of free carriers, via defect-assisted and bimolecular recombination, transfers energy to the lattice, as each recombination event is accompanied by phonon emission, which increase the lattice temperature. 
In this picture, three dynamical degrees of freedom contribute to the permittivity variation: (i) the instantaneous pump intensity which, via TPA, induces an ultrafast increase of the losses, translating into a purely imaginary contribution $\Delta\varepsilon_{\text{TPA}}$; (ii) the population of free carriers, with density $N(t)$, which contributes to a Drude-like modulation $\Delta\varepsilon_{\text{D}}$ with both real and imaginary non-zero components; and (iii) the increased lattice temperature $\Theta(t)$, which modifies the permittivity through a thermo-optic contribution $\Delta\varepsilon_{\text{TO}}$. 
The dynamical evolution of $N(t)$ and $\Theta(t)$ was obtained by solving the Two-Temperature Model (TTM) \cite{Shcherbakov2015,DellaValle2017}, a rate-equation system driven by the pump pulse intensity. 
The total permittivity variation, $\Delta\varepsilon = \Delta\varepsilon_{\text{TPA}}+ \Delta\varepsilon_{\text{D}}+\Delta\varepsilon_{\text{TO}}$, was then analytically computed. 
A more detailed description of the physical mechanisms is provided in \tcr{Methods Section 2}.

To reproduce the astigmatic photoexcitation used in the experiments, we introduced a space- and time-dependent pump intensity $I_\text{pump}(x,t)$. 
The TTM was solved locally for different positions along the metasurface aperture, thereby accounting for the local excitation conditions imposed by the Gaussian pump profile.
This procedure (detailed \tcr{in Methods section 2}) yielded a full spatiotemporal permittivity modulation map $\Delta\varepsilon(x,t)$, which was then used as input for the same full-wave, finite-sized simulations presented in Fig.~\ref{fig1}d-e, adapted to mimic the experimental measurement conditions. 

Specifically, after extracting the complex-valued transmitted probe electric field $\mathbf{E}(x,t)$ at a plane located $\SI{2}{\micro m}$ above the metasurface, we propagated the field through the experimental imaging system using analytical formulas from scalar diffraction theory (details are provided in \tcr{Section S2 of the SI}).
This allowed us to compute the same differential quantity $\Delta I/I_0$ measured by the CMOS camera.
Overall, the simulation provides a full numerical counterpart of the ultrafast imaging experiment, describing the entire process from carrier photoexcitation and nonequilibrium pemittivity modulation to the spatial redistribution of the transmitted probe intensity recorded by the detector.\\

The calculated traces and maps are shown in Fig.~\ref{fig4}b and Fig.~\ref{fig4}d, respectively, and can be directly compared with the experimental results presented in Fig.~\ref{fig4}a and Fig.~\ref{fig4}c. 
The agreement is excellent: the model closely reproduces the measured signals, capturing the double-peaked differential profile characteristic of transient defocusing, its spatial extent, and its sub-picosecond temporal evolution.
At longer pump-probe delays, the simulation also captures the transition towards a predominantly negative signal, consistently with the increasing role of amplitude attenuation over phase-driven intensity redistribution.
This agreement validates the proposed modelling framework and confirms that the observed $\Delta I / I_0$ originates from the resonant conversion of the photoinduced spatial permittivity profile into a transient phase profile that reshapes the transmitted probe wavefront. 

As a control experiment, we performed the same ultrafast imaging measurements on an unpatterned a-Si:H film of comparable thickness, keeping all other experimental conditions unchanged. 
The corresponding experimental and simulated $\Delta I/I_0$ maps are shown in Fig.~\ref{fig4}e-f. 
In this case, the differential intensity signal is dominated by a negative contribution without any peculiar structure along the $x-$axis, indicating an almost homogeneous attenuation of the transmission amplitude and no signature of pump-induced defocusing. 
This comparison shows that spatially structured photoexcitation alone is insufficient to produce the observed wavefront shaping. Instead, the resonant nanostructuring of the metasurface is essential to convert the photoinduced permittivity landscape into a phase-driven redistribution of the transmitted probe field. Together, these results demonstrate ultrafast all-optical wavefront shaping in a metasurface whose optical functionality is not permanently encoded during fabrication,  but is transiently written by the spatiotemporal profile of the excitation light.

\section{Discussion}
\label{sec:discussion}
In summary, we have presented an all-optical strategy for ultrafast wavefront shaping based on a spatiotemporally modulated metasurface. 
In contrast to conventional metalenses, where the target phase profile is permanently encoded in the geometry and spatial arrangement of the meta-atoms, our approach relies on a spatially structured pump pulse to transiently impart an inhomogeneous phase across the metasurface plane.\\

As a proof of concept, we used a strongly astigmatic pump pulse to drive a non-uniform modulation of the optical properties of a one-dimensional a-Si:H nanowire array.
Ultrafast imaging experiments reveal a photoinduced defocusing of the transmitted probe beam in a spectral region selected to maximize phase modulations while minimising amplitude variations, in analogy with the design principle of static metalenses. 
This condition, identified by our numerical model near the isosbestic point of the differential transmission, corresponds to a negative photoinduced phase shift, which acquires a Gaussian spatial profile on a scale of tens of micrometers due to inhomogeneous pump absorption.
By combining full-wave electromagnetic simulations of the whole metasurface aperture with a microscopic description of the non-equilibrium response of Si, our model reproduces the measured spatial and temporal evolution of the differential intensity signal, and traces the observed divergent-lens-like response back to the space-dependent permittivity variations across the meta-atoms.
The comparison with an unpatterned film further highlights the crucial role of nanostructuring.
In the absence of the metasurface resonance, photoexcitation produces a predominant transmission modulation, with no phase-driven redistribution of the probe intensity. \\

Building on these observations, our simulations indicate that the same platform can support additional transient functionalities.
In particular, by operating on the blue wing of the unperturbed resonance, around $\lambda \simeq 666$~nm (see Fig.~\ref{fig1}c) the differential transmission intensity vanishes again, while the differential phase exhibits a positive peak.
This condition, complementary to the one explored above, is predicted to produce a photoinduced converging phase profile, and hence transient focusing of the transmitted probe light on a picosecond timescale, as shown by additional simulations (see SI Section S3).\\

Beyond the specific implementation discussed here, our results establish a general framework for all-optically induced spatial phase modulation in flat optical systems, in which the probe wavelength can be used to select distinct optical functionalities. 
Our strategy can be extended to two-dimensional metasurfaces and more complex spatial profiles of the pump pulse, enabling dynamic wavefront shaping and light structuring. 
More broadly, our work paves the way to reconfigurable flat-optical platforms whose functionality is optically activated and controlled by light on ultrafast timescales.

\section{Methods}

\subsection{1. Ultrafast imaging experiments}
The transient optical response of the metasurface was characterized using a home-built widefield pump-probe microscopy apparatus driven by a \SI{1}{\kilo\hertz} amplified Ti:sapphire laser system (Libra, Coherent) delivering \SI{100}{\femto\second} fundamental pulses at \SI{800}{\nano\meter}. Pump pulses centered at \SI{400}{\nano\meter} were generated via second harmonic generation (SHG) in a $\beta$-barium borate (BBO) crystal. The remaining fraction of the fundamental beam was directed through a near-IR optical parametric amplifier (OPA, center wavelegnth \SI{1240}{\nano\meter}, bandwidth $\sim$\SI{120}{\nano\meter}), and the OPA output was focused into a 4 mm thick YAG crystal to generate a broadband white light continuum for the probe beam. 
A motorized delay stage (PI Instruments) in the pump path controlled the relative temporal delay between pump and probe pulses, while a mechanical chopper (MC200B, Thorlabs) modulated the pump beam at 500 Hz. The pump and probe beams were non-collinearly focused onto the metasurface with the probe at normal incidence and the pump at a small angle (<$7^{\circ}$). A cylindirical lens ($f_{\text{cyl}} = \SI{50}{\milli\meter}$) focused the pump into an astigmatic spot (\SI{45}{\micro\meter} $\times$ \SI{4}{\milli\meter} FWHM), while the probe had a Gaussian profile (\SI{70}{\micro\meter} FWHM), with both beams spatially overlapped at the sample plane. 

The transmitted probe was collected by a 0.9~NA, $100\times$ objective lens (Leica, HC PL Fluotar) and imaged onto a CMOS camera (Pixelink PL-D721MU) through a relay imaging system with achromatic lenses ($f_{\text{L1}} = \SI{75}{\milli\meter}$, $f_{\text{L2}} = \SI{100}{\milli\meter}$). The camera was positioned away from the nominal relay image plane, with an axial offset estimated to be $d=\SI{15}{mm}$, adjusted to increase the contrast of the photoinduced differential signal. This controlled defocusing converts transient phase modulations into measurable spatial intensity redistributions. An interference filter (\SI{10}{\nano\meter} bandwidth) centered at the selected probe wavelength spectrally isolated the signal while rejecting pump scatter. All measurements were performed with TE-polarized probe pulses (electric field parallel to the nanowire orientation) and a pump fluence of \SI{3.5}{\milli\joule\per\centi\meter\squared}.

\subsection{2. Numerical simulations}

\subsubsection{2.1. Periodic model}

To obtain the numerical results presented in Figs.~\ref{fig1}b-\ref{fig1}c, we solved Maxwell's equations in the frequency domain to extract the amplitude and phase of the probe light transmitted by the metasurface, treated as a perfectly periodic, infinite 1D array. 
In defining the computational domain, we exploited the symmetries of the system (i.e., continuous translational invariance along the $y-$axis, discrete translational invariance with periodicity $P$ along the $x-$axis), and built a 2D unit cell in the $xz-$plane, with Floquet periodic boundary conditions at the (vertical) boundaries along the $z-$axis (see Fig.~\ref{fig1}b).
Periodic ports are implemented on the bottom (silica substrate side) and top (air side) boundaries, with the former setting the illumination by a monochromatic plane wave at normal incidence, linearly polarized out of plane (i.e., along the nanowires). 


\subsubsection{2.2. Full finite-scale model}

To extract the simulated $\Delta I/I_0$ (results reported in Fig.~\ref{fig4}), we developed a multi-step model of the whole finite-scale aperture of the metasurface, implementing a full virtual ultrafast imaging measurement according to the algorithm detailed below. 


\paragraph{Pump.} As in the experiments, the simulation considers a spatially inhomogeneous pump pulse impinging on the metasurface from the substrate side, with linear polarisation along the longitudinal direction of the nanowires ($y-$axis).
The small (<$7^{\circ}$) angle of incidence of the pump beam was disregarded, considering that its impact is limited to a minor left-right asymmetry in the experimental modulations, without altering the main mechanism under study.   
The pump pulse intensity is expressed as a function of the $x-$coordinate across the metasurface plane (refer to the reference frame sketched in Fig.~\ref{fig1}) and time $t$ as:
\begin{equation}
    I_{\text{pump}}(x,t) = \Tilde{I}_{\text{p}} \exp\left(-4\log{2} \ \frac{x^2}{I_x^2}\right) \exp\left(-4\log2 \ \frac{t^2}{\tau_{\text{FWHM}}^2}\right),
    \label{eq:pump_intensity}
\end{equation}
with $\Tilde{I}_{\text{p}} \simeq \SI{3.43 E2}{W \per \micro m \squared}$ the peak incident intensity, $I_x \simeq \SI{45}{\micro m}$ the spatial FWHM along $x$ and $\tau_{\text{FWHM}} \simeq \SI{100}{fs}$ the temporal FWHM. These parameters correspond to a simulated incident fluence $F_{\text{sim}} \simeq \SI{3.65}{\milli\joule\per\cm\squared}$, close to the experimental value. We approximate the intensity as uniform along the $y-$direction. \\

\noindent
To model pump absorption across the metasurface, we took into account both one- and two-photon absorption (TPA) processes. We employed the periodic model described above to extract the linear absorption $A_{\text{lin}}$ of a monochromatic plane wave at the pump wavelength $\lambda_{\text{pump}} = \SI{400}{nm}$, obtaining $A_{\text{lin}} \simeq 0.55$. 
As for the TPA, in agreement with refs.~\citenum{Shcherbakov2015,DellaValle2017}, we expressed the nonlinear absorption as $A_{\text{TPA}} = (1-A_{\text{lin}})\left(1-\exp(-\beta_\text{eff}I_{\text{pump}}h)\right)$, with $\beta_{\text{eff}} = \SI{5.2E-10}{m/W}$ a fitted effective TPA coefficient \tcr{(comparable to measured coefficients in a-Si:H thin films of similar thickness\cite{DellaValle2017})}, and $h = \SI{220}{nm}$ the meta-atom thickness. We highlight that $h$ is slightly smaller than the nominal thickness of the fabricated metasurface ($\SI{250}{nm}$), to take into account possible defects originating from the meta-atoms etching (see SI section 1 for further details). Notice that $A_{\text{TPA}}$ is a function of pump intensity and therefore depends on both $x$ and $t$, $A_{\text{TPA}} = A_{\text{TPA}}(x,t)$. 


\paragraph{Local Two-Temperature Model (TTM).}

The TTM is a rate-equation model to describe the nonequilibrium dynamics of photogenerated charge carriers in the material in terms of two energetic variables: (i) $N$, the density (per unit volume) of the hot electron-hole pairs across the semiconductor bands; and (ii) the system lattice temperature $\Theta$.
For a system with a spatially uniform pump intensity $\bar{I}_{\text{pump}}(t)$, the equations of the TTM are given by:
\begin{equation}
    \begin{aligned}
        \frac{\mathrm{d}N}{\mathrm{d}t} &= \Phi(t) - \left(\frac{N}{\tau_{\text{trap}}}+\gamma N^2\right), \\
        \frac{\mathrm{d}\Theta}{\mathrm{d}t} &= \frac{E_{\text{eh}}}{C_{\text{L}}}\left(\frac{N}{\tau_{\text{trap}}}+\gamma N^2\right),
    \end{aligned}
\end{equation}
with $\tau_{\text{trap}} = \SI{30}{ps}$ the trap-assisted recombination timescale\cite{DellaValle2017,Shcherbakov2015}, $\gamma = 7\times 10^{-15} \unit{\metre \tothe{3} \second \tothe{-1}}$ the bimolecular recombination rate \cite{Esser1990}, $C_{\text{L}} = \SI{1.66}{\joule \kelvin \cm \tothe{-3}}$ the lattice heat capacity\cite{DellaValle2017}, and $E_{\text{eh}}$ the energy of the electron-hole pair delivered to the lattice upon each recombination. 
Since the value of $E_{\text{eh}}$ depends on the process of pair creation (it is either the energy of each pump photon $E_{\text{pump}} = 2\pi\hbar/\lambda_{\text{pump}}\simeq \SI{3.1}{\electronvolt}$ for linear absorption, or twice this value for TPA), we considered an averaged value, weighted over the relative probability of linear and nonlinear absorption: $ E_{\text{eh}} = \dfrac{A_{\text{lin}}E_{\text{pump}}+2A_{\text{TPA}}E_{\text{pump}}}{A_{\text{lin}}+A_{\text{TPA}}}$.
Finally, the driving term of the TTM, $\Phi(t)$, comprises both linear and TPA contributions to absorption:
\begin{equation}\label{eq:Phi}
    \Phi(t) = \frac{\lambda_{\text{pump}}}{2\pi \hbar c}\frac{P}{h w}\left(A_{\text{lin}}+\frac{A_{\text{TPA}}}{2} \right)\bar{I}_{\text{pump}}(t).
\end{equation}

To take into account the non-uniform photoexcitation defined by the spatiotemporal intensity $I_{\text{pump}}(x,t)$ of Eq.~\eqref{eq:pump_intensity}, we extended the spatially homogeneous formulation of the TTM to include distinct levels of excitation based on the in-plane position of the meta-atom, thereby mapping the $x-$coordinate of the metasurface.
In particular, we defined a discrete sampling of the $x$-axis, $X_n = \left\{x_i, \ i = 1,\dots, n \right\}$, and a corresponding sampling of the spatially dependent contribution of the incident intensity:
\begin{equation*}
    I_n = \left\{ I_{\text{pump,} i} = \Tilde{I}_{\text{p}} \exp\left(-4\log{2} \ \frac{x_i^2}{I_x^2}\right), \ i = 1,\dots, n \right\}.
\end{equation*}

\noindent
Based on this discrete sampling, we numerically solved the TTM $n$ times, and for each  $i$-th iteration we set a locally homogeneous term, $I_{\text{pump},i} \exp\left(-4\log2 \ \frac{t^2}{\tau_{\text{FWHM}}^2}\right)$, as the driving intensity $\bar{I}_{\text{pump}}(t)$ entering the expression of the source term in Eq.~\eqref{eq:Phi}.
Each of the $n$ local TTMs is solved to determine the temporal dynamics of $N(t)$ and $\Theta(t)$, for the specific $i$-th position considered along the $x-$coordinate. 
From the collection of $n$ spatially homogeneous functions, we then calculated the corresponding local permittivity variation,  creating an interpolation table $\Delta\varepsilon(x,t)$, as detailed below.

\paragraph{Nonlinear permittivity variations.}
To calculate the permittivity variations induced by ultrafast photoexcitation, we considered three main mechanisms: (i) a contribution arising from TPA, (ii) a Drude-like modulation associated with electron and hole plasmas; and (iii) thermo-optics. 
The modulation term due to TPA is an instantaneous nonlinearity producing a dispersion-less change of the absorption coefficient\cite{DellaValle2017} $\Delta\alpha = \beta_{\text{eff}}I_{\text{pump}}$. This yields a permittivity variation given by:
\begin{equation*}
    \Delta\varepsilon_{\text{TPA}} = \mathrm{i}n_{\text{a-Si:H}}(\lambda_{\text{pump}})\frac{\lambda_{\text{pump}}}{2\pi}\Delta\alpha,
\end{equation*}
with $n_{\text{a-Si:H}}(\lambda_{\text{pump}})$ the nanowire real part of the refractive index at the pump wavelength.
An additional contribution to the changes in the material permittivity arises from a Drude-like mechanism caused by the photo-induced electron-hole plasmas\cite{Esser1990,DellaValle2017,Shcherbakov2015}. The real and imaginary parts of $\Delta\varepsilon_{\text{D}}$ read as follows:
\begin{equation}
    \begin{aligned}
        \text{Re}(\Delta\varepsilon_{\text{D}}) &= -\frac{N(t)e^2}{m\varepsilon_0 (4 \pi^2 c^2/\lambda_{\text{probe}}^2+1/\tau_{\text{d}}^2)}, \\
        \text{Im}(\Delta\varepsilon_{\text{D}}) &= - \frac{\lambda_{\text{probe}}\text{Re}(\Delta\varepsilon)}{2\pi c\tau_{\text{d}}},
    \end{aligned}
\end{equation}
with $e$ the electron charge, $m$ the pair effective mass, $m = 0.12m_0$ (with $m_0$ the electron mass), $\varepsilon_0$ the vacuum permittivity, $c$ the speed of light and $\tau_{\text{d}} = \SI{0.8}{fs}$ the damping timescale\cite{DellaValle2017,Shcherbakov2015}. Notice that $\Delta\varepsilon_{\text{D}} = \text{Re}(\Delta\varepsilon_{\text{D}})+\mathrm{i}\text{Im}(\Delta\varepsilon_{\text{D}})$ has a dispersive character, depending on the probe wavelength $\lambda_{\text{probe}}$. 
Finally, we included a phenomenological thermo-optic modulation $\Delta\varepsilon_{\text{TO}}$ that depends on the increase of the lattice temperature $\Delta\Theta(t) = \Theta(t)-\Theta_0$, with $\Theta_0$ the environment temperature:
\begin{equation*}
    \begin{aligned}
        \text{Re}(\Delta\varepsilon_{\text{TO}}) &= 2\left[n_{\text{a-Si:H}}(\lambda_{\text{probe}})\eta - k_{\text{a-Si:H}}(\lambda_{\text{probe}})\zeta\right]\Delta\Theta(t), \\
        \text{Im}(\Delta\varepsilon_{\text{TO}}) &= 2\left[k_{\text{a-Si:H}}(\lambda_{\text{probe}})\eta + n_{\text{a-Si:H}}(\lambda_{\text{probe}})\zeta \right]\Delta\Theta(t).
    \end{aligned}     
\end{equation*}
Here, $n_{\text{a-Si:H}}(\lambda_{\text{probe}})$ and $k_{\text{a-Si:H}}(\lambda_{\text{probe}})$ are the real and imaginary part of the nanowire refractive index evaluated at the probe wavelength. The thermo-optic coefficients are set to $\eta = \dfrac{\mathrm{d}n_{\text{a-Si:H}}}{\mathrm{d}\Theta} = 4.5\times10^{-4}\unit{\kelvin\tothe{-1}}$ (after refs.~\citenum{Fauchet1989,Shcherbakov2015,DellaValle2017}) and $\zeta = \dfrac{\mathrm{d}k_{\text{a-Si:H}}}{\mathrm{d}\Theta} = \dfrac{\kappa\lambda_{\text{probe}}}{(4\pi)}$, with $\kappa = \SI{0.8}{\kelvin\tothe{-1}\cm\tothe{-1}}$ a fitted parameter \tcr{of the same order of the one employed in literature\cite{Shcherbakov2015}}.
The total permittivity variation was then taken as  the sum of these three effects, and expressed as $\Delta\varepsilon = \Delta\varepsilon_{\text{TPA}} + \Delta\varepsilon_{\text{D}}+\Delta\varepsilon_{\text{TO}}$. 
By iterating these calculations for each of the $N$ spatially homogeneous solutions of the TTM, we obtained a collection of $\Delta\varepsilon_i$, mapping the $x-$coordinate across the metasurface, that we collected in an interpolation table $\Delta\varepsilon(x,t)$.
\paragraph{Probe.}
To simulate the optical response of the metasurface upon interaction with the probe beam, both under unperturbed and out-of-equilibrium conditions, we built a finite-sized model of the whole metasurface aperture. 
Since translational invariance along the $y-$direction is enforced, we employed a 2D domain. 
The geometry comprises 739 meta-atoms, and has a total extension of $\SI{300}{\micro m}$ along the $x-$coordinate. 
The probe illumination is defined by imposing an incident Gaussian wave with spatial FWHM $I_{\text{probe}} = \SI{70}{\micro m}$ (in agrement with the experimental estimate) on the substrate bottom boundary, using Scattering Boundary Conditions (SBCs).  
Beyond the top and the lateral boundaries, we implemented Perfectly Matched Layers, with no-incident-field SBCs on their external edges, to absorb any outcoming radiation and avoid spurious reflections.
The metasurface response was then parametrically calculated at different time delays $t$ for the selected probe wavelength $\lambda_\text{probe}$, by expressing the permittivity of the nanowires as $\varepsilon(\lambda_\text{probe}, x, t) =  \varepsilon^0(\lambda_\text{probe}) + \Delta \varepsilon(\lambda_\text{probe}, x, t) $, with $ \varepsilon^0(\lambda_\text{probe})$ the static, unperturbed permittivity of a-Si:H, and $\Delta \varepsilon(\lambda_\text{probe}, x, t)$ the photoinduced modulation term extracted from the interpolation table previously calculated. 
By solving Maxwell's equations in the scattering formalism, we evaluated at each time delay $t$ the complex-valued transmitted electric field $\mathbf{E}(x,t)$ on a plane $\SI{2}{\micro m}$ above the plane of the metasurface, and used it to further derive the differential intensity $\Delta I/I_0 (x,t)$ registered by the experimental camera (details are provided in \tcr{SI section S2}).
For the thin-film simulations (Fig.~\ref{fig4}f), we used the same model by adjusting the geometry accordingly. 
An analogous model was also employed for the proof-of-concept simulations discussed in Fig.~\ref{fig1}e: all the implementation details are as explained above, the only exception being the imposed permittivity variation, which is the one shown in Fig.~\ref{fig1}d.

\section*{Data availability}
All the data supporting this study are available upon request to the corresponding authors.

\section*{Acknowledgements}
M.A., G.~Crotti and M.M. acknowledge financial support from the ERC-StG ULYSSES grant agreement no. 101077181 funded by the European Union. Views and opinions expressed are those of the authors only and do not necessarily reflect those of the European Union or the European Research Council. Neither the European Union nor the granting authority can be held responsible for them. G.~Crotti and G.D.V. acknowledge support from the METAFAST project that received funding from the European Union Horizon 2020 Research and Innovation program under grant agreement no. 899673.
A.S. and G.D.V. acknowledge financial support from the European Union's Horizon Europe research and innovation programme under the Marie Sk\l{}odowska-Curie Action PATHWAYS HORIZON-MSCA-2023-PF-GF grant agreement no. 101153856. 

\section{Author information}

\subsection{Contributions}
These authors contributed equally: Mert Akturk and Giulia Crotti.\\
These authors jointly supervised this work: Giuseppe Della Valle and Margherita Maiuri.

\subsection{Correspondence}
Emails: giuseppe.dellavalle@polimi.it, margherita.maiuri@polimi.it


\clearpage
\bibliography{bibliography}

\end{document}